\documentclass[twocolumn,aps,epsfig,footnote]{revtex4}
\usepackage{amssymb, amsmath, bm, dcolumn, epsf, graphicx, latexsym, mathbbol, slashed}
\usepackage{graphicx}
\usepackage{dcolumn}
\usepackage{bm}
\usepackage{amssymb}
\usepackage{color}
\usepackage{multirow}
\usepackage[colorlinks,linkcolor=blue,citecolor=blue,urlcolor=blue]{hyperref}

 \def\tskip{\setlength{\tskip}{5pt}}
\def\colwidth{\setlength{\colwidth}{3.5in}}

\newcommand{\lsim}{\mathrel{\hbox{\rlap{\lower.55ex\hbox{$\sim$}} \kern-.3em \raise.4ex \hbox{$<$}}}}
\newcommand{\gsim}{\mathrel{\hbox{\rlap{\lower.55ex\hbox{$\sim$}} \kern-.3em \raise.4ex \hbox{$>$}}}}
\newcommand{\beq}{\begin{equation}}
\newcommand{\eeq}{\end{equation}}
\newcommand{\be}{\begin{equation}}
\newcommand{\ee}{\end{equation}}
\newcommand{\bes}{\begin{equation*}}
\newcommand{\ees}{\end{equation*}}
\newcommand{\beqa}{\begin{eqnarray}}
\newcommand{\eeqa}{\end{eqnarray}}
\newcommand{\bea}{\begin{eqnarray}}
\newcommand{\ena}{\end{eqnarray}}

\usepackage{color}

\begin{document}

\title{Model-independent test of the parity symmetry of gravity with gravitational waves}

\author{Wen Zhao$^{1,2}$, Tan Liu$^{1,2}$, Linqing Wen$^{3}$, Tao Zhu$^{4}$, Anzhong Wang$^{5}$, Qian Hu$^{1,2}$, Cong Zhou$^{1,2}$}
\affiliation{$^{1}$ CAS Key Laboratory for Research in Galaxies and Cosmology, Department of Astronomy, University of Science and Technology of China, Hefei 230026, China;  \\
$^{2}$ School of Astronomy and Space Sciences, University of Science and Technology of China, Hefei, 230026, China \\
$^{3}$ Department of Physics, The University of Western Australia, 35 Stirling Hwy, Crawley, Western Australia 6009, Australia\\
$^{4}$ Institute for Theoretical Physics and Cosmology, Zhejiang University of Technology, Hangzhou, Zhejiang 310032, China\\
$^{5}$ GCAP-CASPER, Department of Physics, Baylor University, Waco, Texas 76798-7316, USA}


\begin{abstract}
{Gravitational wave (GW) data can be used to test the parity symmetry of gravity by investigating the difference between left-hand and right-hand circular polarization modes. In this article, we develop a method to decompose the circular polarizations of GWs produced during the inspiralling stage of compact binaries, with the help of stationary phase approximation.
The foremost advantage is that this method is simple, clean, independent of GW waveform, and is applicable to the existing detector network. Applying it to the mock data, we test the parity symmetry of gravity by constraining the velocity birefringence of GWs. If a nearly edge-on binary neutron-stars with observed electromagnetic counterparts at 40 Mpc is detected by the second-generation detector network, one could derive the model-independent test on the parity symmetry in gravity: the lower limit of the  energy scale of parity violation can be constrained within $\mathcal{O}(10^4{\rm eV})$.

}
\end{abstract}


\maketitle

\section {Introduction}

Although Einstein's General Relativity (GR) has been considered to be the most successful theory of gravity since it was proposed, it faces the difficulties in both theoretical side (e.g. singularity, quantization, etc), and observational side (e.g. dark matter, dark energy, etc). Therefore, testing GR in various circumstance is an important topic since its birth \cite{test1,test2,test3,test4}. The discovery of GW compact binary coalescence source GW170817 \cite{gw170817}, and its electromagnetic (EM) counterparts in different frequency bands \cite{gw170817-em}, opens the new window of multi-messenger GW astronomy. This also provides an excellent opportunity to test GR in the strong gravitational fields \cite{gw170817-testGR,gw150914-testGR,gw-nature,gw-white,lorentz}. Numerous works have been carried out, which can be separated into two classes: One is the model-dependent methods, where for a specific theory of gravity, one calculates the GW waveforms, and constrains its deviation from that of GR \cite{bd,1304.3473,1501.07274}. The other is the model-independent methods, which test a specific property of gravity and the results are applicable for a set of gravitational theories \cite{ppe}.



Parity symmetry implies that a directional flipping to the left and right does not change the laws of physics. It is well known that nature is parity violating. Since the first discovery of parity violation in weak interactions \cite{Lee-Yang}, the experimental tests become more necessary in the other interactions, including gravity. In most previous works, test of parity symmetry in gravity has focused on the Chern-Simon modified gravity (see for instance the review article \cite{CS-review}). Many parity-violating (PV) gravities with different action forms have been proposed for different motivations \cite{unify,qiao,gao,horava,wang-review,tele}. If the parity symmetry is violated, it is expected that a GW behaves asymmetrically in its two circular polarization modes \cite{effective-field,zhao2019}. The observable effects include the ``amplitude birefringence"\cite{0410230,cs1,cs2,cs3,cs4} and the ``velocity birefringence" \cite{zhu_effects_2013,1809,tele,soda,wang_polarizing_2012,gao,wang-review,wang2020}. Although Chern-Simons gravity causes only the amplitude birefringence of GWs, both effects occur in most PV gravities (see Appendix \ref{appendixA}). Therefore, reconstructing circular polarizations from observed GW signals is crucial.

{\color{black}In this article, for the general PV gravities with velocity birefringence effect, we focus on the GW signal emitted by the coalescence of compact binaries, and calculate the arrival times of the left-hand and right-hand polarization modes. We find that the arrival time difference between two modes directly depends on the energy scale of parity violation in gravity. In order to measure the arrival time difference,
we develop a waveform-independent method to reconstruct the circular polarizations of GWs with the help of stationary phase approximation (SPA), and measure the arrival times these two modes from observed data. By mocking the potential observation of the upcoming second-generation GW detector network, we test the reliability of the method, and apply it to constrain model-independently the velocity birefringence of GWs. The foremost advantage is that this method is simple (without tedious calculation), clean (with least assumption), independent of GW waveform, and is applicable to the existing detector network.}

\section{Circular polarizations of GWs}

\subsection{The decomposition method}
In general, it is convenient to describe GWs with {\emph{complex}} oscillating functions $h_{\rm s}(t)$ \cite{mtw}. The detector response is a linear combination of the {\emph{real}} part of two wave polarizations \cite{300}, {\emph{i.e.}}
$d_I(t+\tau_I)=\sum_{s=+,\times}F_I^{s} h_{s}^{\mathfrak R}(t)+n_I(t+\tau_I)$, where $I=1,2,3,\cdot\cdot\cdot$ labels the $I$-th detector, and the superscript $\mathfrak R$ denotes the real part of the corresponding quantity in this paper. $\tau_I$ is the relative time delay with respect to a reference time $t$ taken at the center of the Earth. $F_I^{s}$ are the detector's beam-pattern functions. For a network with $I\ge 2$, in principle, an unbiased estimator for polarization mode $h_{s}^{\mathfrak R}(t)$ can be solved directly from the data $d_{I}$ by introducing the Moore-Penrose psudo-inverse matrix $A$, which is composed of the detectors' antenna beam-pattern functions $F_I^{\rm s}$ \cite{wen2007,hayama2012}. For a given source direction, the time-delay corrected data from a network of $N_d$ GW detectors can be written in the frequency domains as
$\tilde{\rm {\bf d}}={A} {\rm \tilde{\bf h}^{\mathfrak R}} + \tilde{\rm {\bf n}}$,  ${\rm \tilde{\bf h}^{\mathfrak R}} = \left(\tilde{h}_{+}^{\mathfrak{R}},  \tilde{h}_{\times}^{\mathfrak{R}} \right)^{T}$, and ${A}$ is the response matrix of the detector network at each frequency defined by Eq.(4) in \cite{wen2007},
\bea \nonumber
{\rm \tilde{\bf d}}&=&\left(\frac{\tilde{d}_{1}}{\sigma_{1}},\cdot\cdot\cdot,\frac{\tilde{d}_{N_d }}{\sigma_{N_d }}\right)^{T},
{\rm \tilde{\bf n}}=\left(\frac{\tilde{n}_{1}}{\sigma_{1}},\cdot\cdot\cdot,\frac{\tilde{n}_{N_d }}{\sigma_{N_d }}\right)^{T},
\ena
where $\sigma^2_{i}$ is the noise variance of $i$-th detector at the corresponding frequency bin. Throughout this paper, {\emph{tilde}} denotes the quantity in the frequency domain. The estimator for the vector $\{{\rm \tilde{\bf h}^{\mathfrak R}}\}$ is given by the Moore-Penrose inverse \cite{wen2007}, {\emph{i.e.}}
\bea\label{eq11}
{\rm \tilde{\bf h}^{\mathfrak R}}=(A^{\dagger}A)^{-1} A^{\dagger} {\rm \tilde{\bf d}}.
 \ena
By the inverse Fourier transformation, the estimators for the GW signal in time domain $h_{s}^{\mathfrak R}(t)$ can be derived.

The circular polarizations of GWs are defined by the {\emph{complex}} $h_{s}(t)$ reconstructed from the observables $h_{s}^{\mathfrak R}(t)$. It is direct to prove that this reconstruction is achievable, if considering the SPA, which is applicable for the GWs produced during the inspiralling stage of coalescing compact binaries. Consider $h_+$ as an example, it can be shown that the Fourier components of $h_{+}(t)$ and $h_{+}^{\mathfrak R}(t)$ satisfy the following relations (see the details in Appendix \ref{appendixB}),
\bea
\tilde{h}_{+}=\left\{
 {\begin{array}{l}
 2\tilde{h}_{+}^{\mathfrak{R}},~f>0,  \\
 0, ~~~~~f<0,
 \end{array}
 }
 \right.
\ena
that is, one could derive $\tilde{h}_{+}$ from $\tilde{h}_{+}^{\mathfrak{R}}$. This plays a crucial role in the method proposed in this article, that reconstructs the GW circular polarizations from observation. However, we should emphasize that, this conclusion is applicable only for the inspiraling stage of GW events, where SPA is appropriate. The basic reason for this reconstruction to work is that, in SPA, in each small time span, the amplitude of a GW can be considered as a constant, it is therefore possible to derive both amplitude and phase information of GWs, instead of the combination of them in a general case.
Equivalently, GWs can be decomposed as left-hand (L) and right-hand (R) circular polarizations,  which are defined as $\tilde{h}_{\rm R/L}=({\tilde{h}_{+}\pm i\tilde{h}_{\times}})/{\sqrt{2}}$ \cite{mtw}.
Using the inverse Fourier transformation, we can also obtain the time-domain function $h_{\rm R}(t)$ and $h_{\rm L}(t)$, which are both complex functions.

In this decomposition, the antenna beam-pattern functions of the $I$-th detectors $F^{s}_{I}$ should be known in advance, which depend on the sky direction of the GW signal (RA,Dec), and the polarization angle $\psi_s$ \cite{pattern-function}. The former is assumed to be obtained from its EM counterparts, while the latter remains unknown without a template fitting. On the other hand, a change of $\psi_s$ corresponds to the rotation of the celestial sphere frame coordinate system along $z$-axis \cite{pattern-function}. Considering a rotation of coordinate system along $z$-axis by an angle $\psi$, the left-hand and right-hand modes in the new coordinate system are given by $h'_{\rm{L}/{\rm R}}= h_{{\rm L}/{\rm R}}e^{\pm2\psi}$ \cite{mtw}, which means that this rotation of coordinate system is completely equivalent to a change of the phase of $h_{\rm R}$ and $h_{\rm L}$ with the same value but opposite signs. For this reason, if we care only about the phase evolution, instead of their exact values of phases, the polarization angle $\psi_s$ can be arbitrarily chosen in the reconstruction, as confirmed in the simulation analysis.

\begin{figure}[htbp]
\small
\centering
\includegraphics[width=9cm]{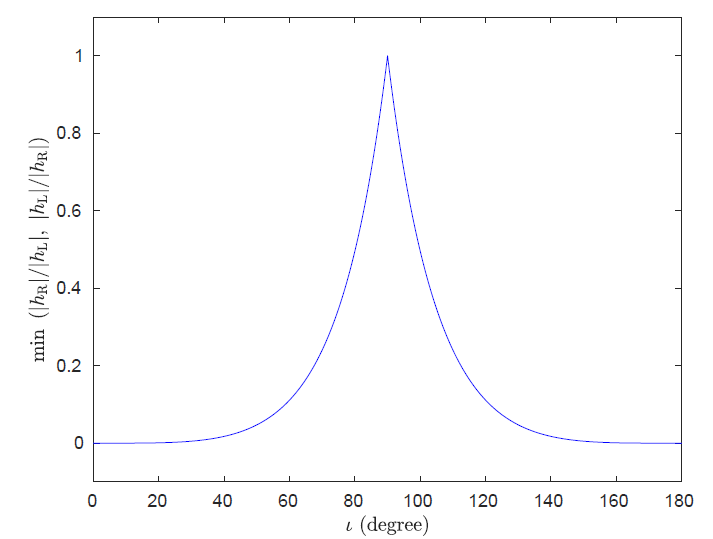}
\caption{The ratio between the amplitude of left-hand polarization and that of right-hand polarization as a function of the inclination angle $\iota$. }
\label{fig5}
\end{figure}

{\color{black}Note that for GWs emitted by the coalescence of compact binaries, the amplitude ratio of left-hand and right-hand polarization modes is determined by the inclination angle $\iota$ of GW event. For the face-on sources with $\iota=0^{\circ}$, or $180^{\circ}$, GW is circularly polarized, {\emph{i.e.}} only left-hand or right-hand polarization exists. While for the edge-on sources with $\iota=90^{\circ}$, the amplitude of two circular polarizations are same. In GR, the amplitude ratio is given by $|h_{\rm L}|/|h_{\rm R}|=[(1+\cos\iota)/(1-\cos\iota)]^2$, which is clearly presented in Figure \ref{fig5}. The condition $|h_{\rm L}|/|h_{\rm R}|\in(1/3,3)$ requires that $\iota\in(74^{\circ},106^{\circ})$. Considering this fact and assuming the deviation from GR is small, we find that for the method proposed in this paper, only the nearly edge-on GW events are suitable for testing the parity symmetry in gravity \cite{ft0}.} 

\subsection{Applying to the simulated data}

We test the reliability of this decomposition method by simulations. In our analysis, GR is considered as the fiducial theory of gravity. We consider a GW signal produced by the coalescence of binary neutron-stars (BNSs), which are accompanied by observable EM emissions in various frequency bands. Without lose of generality, we assume a GW event produced by the coalescence of BNS located at (RA=0, Dec=0). Similar to GW170817, the luminosity distance is adopted as $d=40$ Mpc. Both masses of NSs are chosen as $1.4{\rm M}_{\odot}$ and their tidal deformability parameter are assumed to be 425, referring to Figure 5 of Ref. \cite{gw170817}. For the inclination angle $\iota$, we consider both the face-on case with $\iota=0^{\circ}$ and edge-on case with $\iota=90^{\circ}$. Since we are only interested in GWs in the inspiraling stage, TaylorF2 model is used to calculate GW waveforms in time domian. 

To mimic the realistic case, we consider the detector network including AdvLIGO \cite{aligo}, AdvVirgo \cite{avirgo}, and KAGRA \cite{kagra}, as their designed sensitivities.  The gaussian noise for each detector is randomly generated using the PyCBC package \cite{pycbc}. Mock data set for each detector is a linear combination of signal and noise.

\begin{figure}[htbp]
\small
\centering
\includegraphics[width=8.5cm]{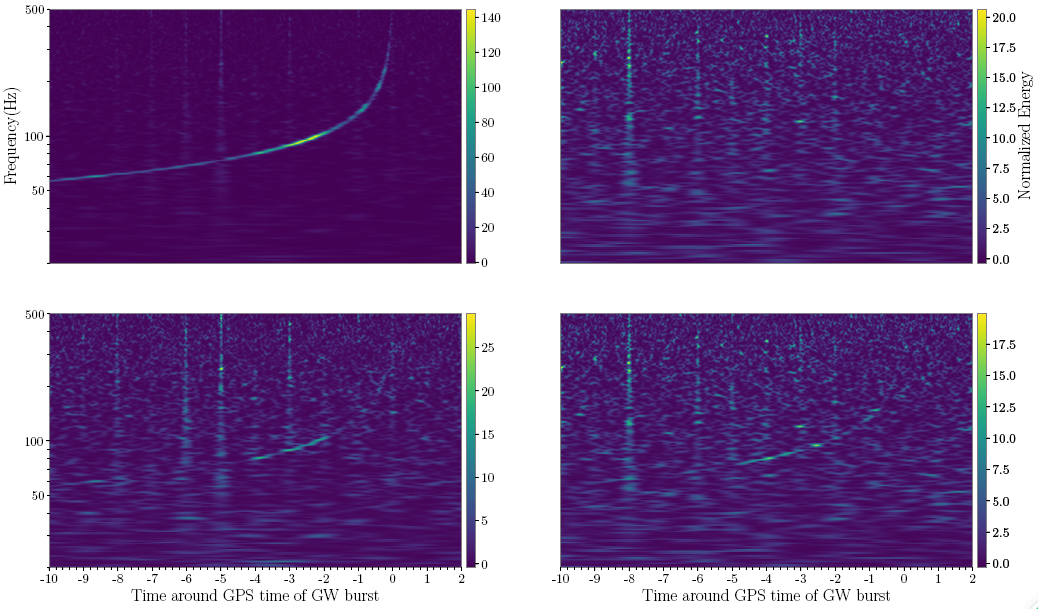}
\caption{Time-frequency representations of left-hand (left panels) and right-hand (right panels) polarization signals reconstructed from the simulated GW data of signal $+$ noise for the second-generation detector network. Upper panels show the results of a face-on GW event, and the lower panels show those of an edge-on event.}
\label{fig1}
\end{figure}

Following the procedure described above, we reconstruct the circular polarizations of GWs for two extreme cases of the inclination angles. Similar to LIGO/Virgo collaboration in \cite{gw150914,gw170817}, we present the results in time-frequency representations of the reconstructed GW signals \cite{gw150914,gw170817}, as shown in Figure \ref{fig1}. We have applied the $Q$-transformation to the reconstructed complex strain serials, and optimized the $Q$ value in the range of (100,110)Hz to maximize signal-to-noise ratio (SNR) in the diagram. The $Q$ transform is a modification of the standard short time Fourier transform in which the analysis window duration varies inversely with frequency \cite{q-transformation}. For the case with face-on burst, we find only the left-hand polarization mode, manifests itself as the significant signal in the upper left panel in Figure \ref{fig1}. This result is anticipated, since in GR, the amplitude of left-hand mode is $h_{\rm L}\propto (1+\cos\iota)^2$ and that of right-hand mode is $h_{\rm R}\propto (1-\cos\iota)^2$ \cite{300}. Therefore, in the extreme case with $\iota=0$, only the left-hand mode exists. While for the case with edge-on source, in the lower panels of Figure \ref{fig1} we observe both polarizations with similar power as expected. 

\section{Testing the chirality of gravity}
\subsection{Arrival time difference and the parity violation in gravity}
In the general PV gravity, GWs propagating in the flat Robertson-Walker universe satisfy \cite{effective-field,zhao2019},
\bea
\tilde{h}_{\rm A}''+(2+\nu_{\rm A})\mathcal{H}\tilde{h}'_{\rm A}+(1+\mu_{\rm A})k^2 \tilde{h}_{\rm A}=0,~({\rm A}={\rm R},{\rm L}),
\ena
where {\emph{prime}} denotes the derivative with respect the conformal time $\eta$, and $\mathcal{H}\equiv a'/a$. $a$ is the scale factor, and the present value is adopted as $a_0=1$. $k$ is the wave number, which relates to the GW frequency by $f=k/2\pi a$. Nonzero $\nu_{\rm A}$ and $\mu_{\rm A}$ represent the deviation from GR \cite{grishchuk,liu,weinberg}. Term $\nu_{\rm A}$ determines the amplitude evolution of GWs, and the term $\mu_{\rm A}$ represents velocity of GWs. The case with $\nu_{\rm R}\neq \nu_{\rm L}$ means different dampings of left-hand and right-hand polarizations, which is the effect of ``amplitude birefringence" \cite{0410230}. For an individual frequency, the effect of amplitude birefringence completely degenerates with the value of inclination angle \cite{CS-review}. Therefore, in principle, this effect can be tested by comparing the amplitude ratio of left-hand and right-hand polarizations among different frequencies, which is left as a future work.

The case with $\mu_{\rm R}=\mu_{\rm L} \neq 0$ represents the violation of Lorentz symmetry in gravity, which has been tightly constrained by comparing the arrival times of GW170817 and GRB170817a \cite{gw170817-speed}. In this paper, we consider only the PV case. $\mu_{\rm R}\neq \mu_{\rm L}$ means the velocities of GW polarizations are different, that is, there exists the ``velocity birefringence". In the general PV gravities, including the ghost-free PV theories of gravity \cite{unify,1809,gao,zhao2019}, Horava-Lifshitz gravity \cite{horava,zhao2019}, PV extension of the symmetric teleparallel equivalent of GR theory \cite{tele,zhao2019} etc, $\mu_{\rm A}$ can be parameterized as
$\mu_{\rm A}=\alpha \rho_{\rm A} (k/aM_{\rm PV})^{\beta}$, with $\rho_{\rm R}=1$ and $\rho_{\rm L}=-1$, $M_{\rm PV}$ is the energy-scale of the theory, $\alpha$ and $\beta$ are the coefficients, which depend on the theory of modified gravity (see the details in Appendix \ref{appendixA}). In the local universe, we can ignore the time-dependence, and treat $\alpha$ as a constant, which is absorbed by the definition of $M_{\rm PV}$ as discussed below. Since the measurements of GWs using laser interferometers are not sensitive to the GW amplitude, in this article we restrict attention to the effect of velocity birefringence. The parametrization of $\mu_{\rm A}$ can be equivalently written as the modified dispersion relation \cite{zhao2019}
\begin{equation}\label{dispersion}
\omega_{\rm A}^2(k)= k^2[1 + {\rm sgn}(\alpha)\rho_{\rm A} (k/aM_{\rm PV})^{\beta}],
\end{equation}
which follows the group velocity of GWs, {\emph{i.e.}}
\bea
v_{\rm A}/c=1-{\rm sgn}(\alpha)(1/2)\rho_{\rm A}(k/aM_{\rm PV})^{\beta}.
\ena
Note that since the sign of $v_{\rm A}/c-1$ is determined by $\rho_{\rm A}$, $\rho_{\rm R}$ and $\rho_{\rm L}$ have opposite signs. If one polarization mode is superluminal, then the other is subluminal.

For a given GW signal at redshift $z$ emitting both left and right circular polarizations, different propagation velocities will result in the difference in signal arrival times, which is given by 
\bea\nonumber
t_{\rm R-L}=(1+z)\Delta t_e +{\rm sgn}(\alpha){(\rho_{\rm R} k_{\rm R}^{\beta}-\rho_{\rm{L}}k_{\rm{L}}^{\beta})T_{\beta}}/{2M_{\rm PV}^{\beta}},
\ena
where $T_{\beta}\equiv \int_{t_e}^{t_0}a^{-\beta-1}dt$. $\Delta t_e$ is the emitting time difference of the two modes. For the comparison of GW with their EM counterparts, the uncertainty of $\Delta t_e$ is the main problem for the GW velocity measurement, which strongly depends on the theoretical models of GW and EM bursts. For instance, LIGO/Virgo collaboration has assumed that the emitting time difference between GW signal of GW170817 and EM signal of GRB170817a is smaller than 10$s$ predicted in some models \cite{gw170817-speed}, and use this to constrain the Lorentz symmetry of gravity \cite{lorentz2}. Therefore, the constraint of GW velocity by comparing the arrival times of GW and EM signals is model-dependent. Fortunately, we do not need to make such assumptions in our method.  Since circular polarizations of GWs are the spin $2$ modes, independent of theory of gravity, both modes are produced by the instantaneous acceleration of mass quadrupole of the systems. For this reason, for a fixed wavenumber, {\emph{i.e.}} $k_{\rm R}=k_{\rm L}$, we have $\Delta t_e=0$, and arrival time deference becomes
\begin{equation}\label{z1}
|t_{\rm R-L}|
=\left({k}/{M_{\rm PV}}\right)^{\beta}T_{\beta}.
\end{equation}
{\color{black}As one of the main results in this article, this formula gives a direct relation between the arrival time difference and the energy scale of parity violation $M_{\rm PV}$. The non-zero measurement of $|t_{\rm R-L}|$ will imply the detection of velocity birefringence effect of GWs, which reflects the parity violation in gravity. On the other hand, if the deviation from zero of $|t_{\rm R-L}|$ cannot be detected, we can place a bound on the energy scale $M_{\rm PV}$, below which the velocity birefringence effect of GWs does not exist. 
}


\subsection{Measurement of arrival times}

\begin{figure}[htbp]
\small
\centering
\includegraphics[width=9cm]{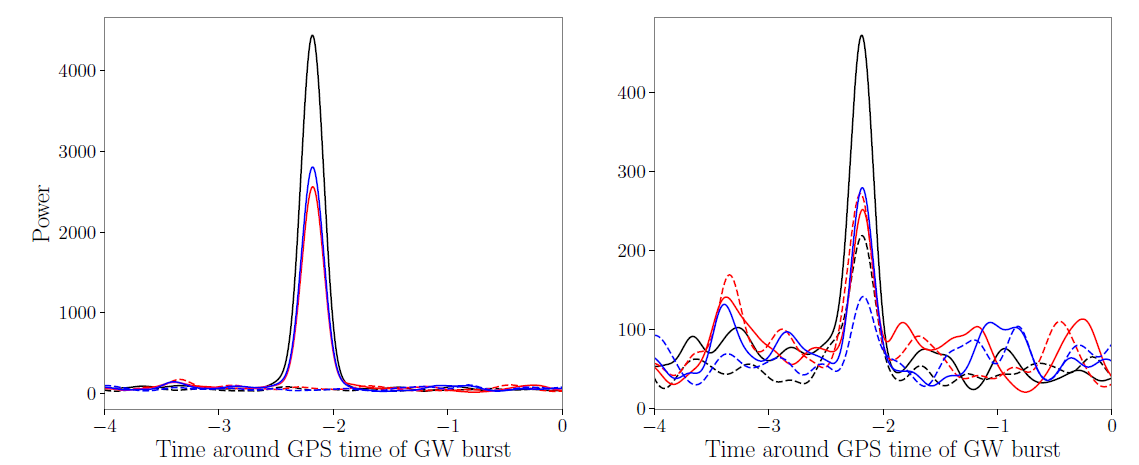}
\caption{Arrival times of left-hand (solid lines) and right-hand polarization (dashed lines) signals at (90,110)Hz. Left panel shows the results of face-on GW events, and
right panel shows that of edge-on events. 
The black lines show the results of the GW event as in Figure \ref{fig1}, and
the red and blue lines show the results of other two samples respectively. Note that, in this figure, the vertical axis stands for the strength of power in Figure \ref{fig1}.}
\label{fig3}
\end{figure}

In realistic observations, the differences of arrival times of two GW polarizations are observable, as long as the left-hand and right-hand polarization can be reconstructed. For the cases with low noise level, {\it e.g.} for the third-generation detector network, the SNR for each frequency channel can be large enough, and we can read out the arrival times of both modes from time-frequency representation, and calculate directly the time difference $t_{\rm R-L}$. However, in the case with the second-generation detectors, the signal for individual frequency channel is too noisy, we have to combine the data within a frequency bin $(f_{lower}, f_{upper})$ to amplify the SNR. To realize it, for each polarization mode, we calibrate all frequencies in the bands to compensate the emitting-time difference for different frequencies. To the lowest Newtonian order, the emitting time of GW at frequency $f$ is given by \cite{spa}
\bea\label{eq10}
 t_c-t= 2.18s \left({1.21{\rm M}_{\odot}}/{\mathcal{M}_c}\right)^{5/3}\left({100{\rm Hz}}/{f}\right)^{8/3},
 \ena
where $t_c$ is the time at which $f$ becomes infinity and $\mathcal{M}_c$ is the chirp mass of the binary. Utilizing this formula, for the simulated data, we calibrate all the frequency bands in the span $(90,110)$Hz to $f=100$Hz, and superimpose them to obtain the arrival times. The results are presented with black lines in Figure \ref{fig3}. From this figure, we obtain the arrival times of left-hand mode $t_{\rm L}=-2.186^{+0.085}_{-0.088}s$ and right-hand mode $t_{\rm R}=-2.182^{+0.088}_{-0.092}s$, by measuring the FWHM of the signal in Figure 2. The difference between them is derived directly $t_{\rm R-L}=0.004^{+0.122}_{-0.127}s$. We observe that, the uncertainty of arrival time difference can be achieved at the level of $\mathcal{O}(0.1s)$ for this particular GW source configuration. Note that, in our analysis, the uncertainties of $t_{\rm R}$ and $t_{\rm L}$ depend not only on the errors of arrival times in GW measurement, but also on the time resolution of the time-frequency representation in Figure \ref{fig1}. Therefore, the uncertainty of time difference $t_{\rm R-L}$ derived above might not be the optimal result, which is expected to be significantly reduced if a better way can be used to read out the arriving times of GW signals. {\color{black}We stress that, although these results are derived from an individual simulation, the stability of the conclusions is confirmed by repeating the analysis above but adopting different realizations. For instance, in Figure \ref{fig3}, we plot the results of the other two cases with red and blue lines, where we consider the same GW events and adopt the different realizations of detector noises. We find that, although the amplitudes of two modes are sensitive to the noise realizations, the uncertainties of their time difference are quite stable. All three samples follow the similar result of $|t_{\rm R-L}|\lesssim 0.12s$. Similarly, considering the GW events with different sky positions, polarization angle, or different neutron-star masses, we find the values of $|t_{\rm R-L}|$ have no significant change. This is caused by the fact that, in this method, the uncertainties of $t_{\rm R}$ and $t_{\rm L}$ are dominated by the time resolution of the time-frequency representation. }

\subsection{Constraining the parity symmetry of gravity}

The value of $|t_{\rm R-L}|$ can be directly translated to the constraint on PV gravity. The analysis above indicates that if a nearly face-on BNSs at 40 Mpc observed by second generation network, $|t_{\rm R-L}|\lesssim 0.12s$ at frequency $f\sim100$Hz is expected to be obtained. The first constraint is for the velocity difference between polarization modes. Using the relation $v_{\rm R}/v_{\rm L}-1=t_{\rm R-L}/ d$, and $d=40$ Mpc, we found that $|v_{\rm R}/v_{\rm L}-1|<3.1\times 10^{-17}$. The constraint on the PV energy scale $M_{\rm PV}$ is also expected to be obtained. From the relation in Eq.(\ref{z1}) we have $k/M_{\rm PV} =(|t_{\rm R-L}| /T_{\beta})^{1/\beta}$. As in general, we consider the case with $\beta=1$, {\emph{i.e.}} the gravity includes the lowest-order PV terms in GWs. Considering the Lambda Cold Dark Matter ($\Lambda$CDM) model with cosmological parameters $H_0=70.0$km/$s$/Mpc, $\Omega_{\Lambda}=0.7$, $\Omega_{m}=0.3$ \cite{weinberg}, we have $T_{\beta}=4.1\times 10^{15}s$, it follows that $M_{\rm PV}>1.4\times 10^{4}$eV. 

\section{GW170817}

At this writing, GW170817 is the unique GW event with observed EM counterparts. Therefore, the reconstruction of two circular polarizations for this event is carried out in Appendix \ref{appendixC}. However, for the event GW170817, observation gives that constraint $\iota\ge152^{\circ}$ \cite{gw170817}, which follows that $|h_{\rm L}|/|h_{\rm R}|<0.004$. So, for this event, the right-hand mode is completely dominant, which is also proved in our analysis. For this reason, the method introduced above cannot be applied in GW170817.

\section{Conclusions}

The advent of multi-messenger GW astronomy opens a new window to test the characteristics of gravity in strong gravitational-field. {\color{black}In this article, we develop a method to reconstruct the circular polarization modes of GWs, emitted during the inspiralling stage of compact binaries, with the help of the source information obtained from the observations of EM counterparts. By simulating the mock GW data, we test the reliability of the decomposition method for various cases.}

By measuring the arrival time difference of left-hand and right-hand polarizations of a GW, one can model-independently test the parity symmetry of gravity. {\color{black}Since the GW observation is more sensitive to the arrival time rather than the amplitude, as an example of application, we mainly focus on the velocity birefringence effect of GW in this paper. For the GW signals emitted by the coalescence of compact binaries, we calculate the arrival time difference of two circular polarization modes in the general PV with velocity birefringence effect, and find it directly relates to the energy scale $M_{\rm PV}$ of parity violation in gravity. Therefore, the measurement of arrival time difference can be used to place the bound of $M_{\rm PV}$.} We test the velocity birefringence of GWs by means of mock data, and find that if a nearly edge-on BNS event at 40 Mpc is observed by the second generation GW detector network, the arrival-time difference can be constrained at the accuracy of $\mathcal{O}(0.1s)$ for the GWs at $f\sim 100$Hz. It follows that the fractional velocity difference of two modes can be constrained at the level of $\mathcal{O}(10^{-17})$. For the general theories of gravity with lowest-order PV terms (except for the Chern-Simons modified gravity, in which the velocity birefringence effect does not exist as mentioned above), this result implies the expected constraint on the energy scale of parity violating $M_{\rm PV} \gtrsim \mathcal{O}(10^{4})$ eV. {\color{black}We emphasize that, the method introduced in this paper can be applied for any GW observations by ground-based detectors or space-borne detectors, as long as the GW signal is emitted by the nearly face-on coalescence of compact binaries with detected EM counterparts.} It is instructive to compare this constraint with the existing constraints of PV gravity \cite{ft1}. We find that this constraint is 17 orders better than the existing constraint in Solar System, which is $M_{\rm PV}\gtrsim 10^{-13}$eV \cite{cs-solar}, and 14 orders better than those derived from binary pulsars \cite{cs-pulsar} or amplitude birefringence of GW \cite{cs3}, which are $M_{\rm PV}\gtrsim 10^{-10}$eV. In \cite{1809}, the authors obtained a constraint of $M_{\rm PV}\gtrsim 10$eV by comparing the arrival-time difference between GW170817 and its EM counterpart GRB170817a, which is 3 orders worse than that derived in this work.



\begin{acknowledgments}
We appreciate the helpful discussions with Mingzhe Li, Weiwei Zhu, Zhoujian Cao and Yifan Wang.
W.Z. is supported by NSFC Grants No. 11773028, No. 11633001, No. 11653002, No. 11421303, No. 11903030, the Fundamental Research Funds for the Central Universities, and the Strategic Priority Research Program of the Chinese Academy of Sciences Grant No. XDB23010200. L.W.'s research is supported in part by Australian Research Council. T.Z. and A.W. are supported by NSFC Grants No. 11675143, No. 11675145, No. 11375153.
\end{acknowledgments}

\appendix

\section{Gravitational wave in the parity-violating gravities \label{appendixA}}

{\color{black} In this Appendix, we will summarize the GW waveforms in various PV gravities. Although the main results have been derived in our previous work \cite{zhao2019}, the brief introduction will be helpful to understand the application of the constraint derived in this paper for various PV gravities.}

In the flat Friedmann-Robertson-Walker universe, GW is the tensor perturbation of the metric, {\emph{i.e.}}
\begin{equation}
ds^2=a^2(\eta)\left[-d\eta^2+(\delta_{ij}+h_{ij})d\chi^i d\chi^j\right],
\end{equation}
where $a(\eta)$ is the conformal scale factor, $\eta$ is the conformal time and $\chi^i$ is the comving coordinates. The quantity $h_{ij}$ stands for the GW perturbation, which we take to be transverse and traceless gauge, $\delta^{ij}h_{ij}=0$ and $\partial_i h^{ij}=0$.

The equation of motion of GW is determined by the tensor quadratic action, which reads
\bea
S^{(2)}=\frac{1}{16\pi G} \int dt d^3x~a^3\left[\mathcal{L}^{(2)}_{\rm GR}+\mathcal{L}^{(2)}_{\rm PV}+\mathcal{L}^{(2)}_{\rm other}\right],
\ena
where $\mathcal{L}^{(2)}_{\rm GR}$
is the standard Lagrangian obtained from the Einstein-Hilbert term $R$. In the viewpoint of effective fields \cite{effective-field}, the first possible corrections to the tensor mode come from terms with three derivatives. In the unitary gauge, the standard quadratic action is modified by the addition of \cite{effective-field}
\[
\mathcal{L}^{(2)}_{\rm PV}=\frac{1}{4} \left[\frac{c_1(t)}{aM_{\rm PV}}\epsilon^{ijk} \dot{h}_{il}\partial_{j}\dot{h}_{kl}+\frac{c_2(t)}{a^3M_{\rm PV}}\epsilon^{ijk}\partial^2 h_{il}\partial_{j} h_{kl}\right],
\]
where a {\emph{dot}} denotes the derivative with respect to the cosmic time $t$, $\epsilon^{ijk}$ is the antisymmetric symbol, $c_1$ and $c_2$ are dimensionless coefficients, which could depend on time, and $M_{\rm PV}$ is the scale that suppresses these higher dimension operators.

Decomposing the GW in the circular polarization basis, in the frequency domain the equation of motion can be written as \cite{1809}
\bea\label{A8}
\tilde{h}_{\rm A}''+(2+\nu_{\rm A})\mathcal{H}\tilde{h}'_{\rm A}+(1+\mu_{\rm A})k^2 \tilde{h}_{\rm A}=0,
\ena
where ${\rm A=R }$ or ${\rm L}$, standing for the right-hand or left-hand polarization mode respectively, and
\bea
\nu_{\rm A}&=&\rho_{\rm A}(k/aM_{\rm PV})(c_1-c_1'\mathcal{H}^{-1}), \label{A8a} \\
\mu_{\rm A}&=&\rho_{\rm A}(k/aM_{\rm PV})(c_1-c_2). \label{A8b}
\ena
Note that, Eq.(\ref{A8}) is the unifying description for low-energy effective description of generic parity-violating GWs. To our knowledge, all the known parity-violating theories of gravity in the literature can be casted into this form \cite{zhao2019}.

{\emph{Chern-Simons (CS) modified gravity}} with Pontryagin term coupled with a scalar field corresponds to the case with $c_1=c_2$, {\emph{i.e.}} $\nu_{\rm A}\propto \rho_{\rm A}(k/aM_{\rm PV})$ and $\mu_{\rm A}=0$ \cite{0410230}. Due to the disappearing of $\mu_{\rm A}$ term, only the amplitude birefringence effect exists in CS modified gravity. 

{\emph{Ghost-free parity-violating theories of gravity}} have recently been explored. One of the theories has the Lagrangian $\mathcal{L}_{\rm PV}$ (see Eqs.(3.1), (3.2) and (3.4) in \cite{unify}), which includes the scalar field and its first derivatives. GW in this theory corresponds to the case with nonzero functions of $c_1\neq c_2$ \cite{1809,qiao}.

{\emph{Another ghost-free parity-violating theory}} contains second derivatives of the scalar field, and the Lagrangian $\mathcal{L}_{\rm PV}$ is given by Eqs.(3.12), (3.13), (3.14) and (3.18) in \cite{unify}, which corresponds to the case with $c_1\neq 0$, $c_2=0$ \cite{1809,qiao}. 

{\emph{Horava-Lifshitz (HL) gravity}} is power-counting renormalizable theory because of the presence of high-order spatial derivative operators \cite{horava,wang-review}.  
The theory with terms of third-order spatial derivatives, {\emph{i.e.}}  three-dimensional gravitational CS term, corresponds to $c_1=0$ and $c_2\neq 0$, which is equivalent to $\nu_{\rm A}=0$, $\mu_{\rm A}\propto \rho_{\rm A}(k/aM_{\rm PV})$ \cite{wang_polarizing_2012,zhu_effects_2013}.

{\emph{Alternative version of HL theory}} is also investigated in literature \cite{soda,wang_polarizing_2012,zhu_effects_2013}, which contains only the fifth-order spatial derivative operators in the parity-violating terms.
In this theory, the equation of motion for GWs is given by Eq.(\ref{A8}) with $\nu_{\rm A}=0$, $\mu_{\rm A}\propto \rho_{\rm A}(k/aM_{\rm PV})^3$ \cite{soda,wang_polarizing_2012,zhu_effects_2013}. {\color{black}As expected, we find these terms would be more suppressed by the energy scale $M_{\rm PV}$ in the view of effective field theories.}

{\emph{Parity-violating extension of the symmetric teleparallel equivalent of GR theory}} is a non-Riemannian formulation of gravity \cite{tele,zhao2019}. 
Considering the three-derivative parity-violating terms, the field equation of GW corresponds to $\nu_{\rm A}\propto\rho_{\rm A}(k/aM_{\rm PV})$ and $\mu_{\rm A}\propto \rho_{\rm A}(k/aM_{\rm PV})$. 

{\emph{Another parity-violating theory}} is to consider only the fifth-order derivative operator $\mathcal{L}^{(2)}_{\rm PV}\propto M_{\rm PV}^{-3}\epsilon^{ijk} \dot{h}_{il}\partial^2\partial_j\dot{h}_{kl}$ \cite{1809}, which corresponds to the equation of motion of Eq.(\ref{A8}) with $\nu_{\rm A} \propto \rho_{\rm A}(k/aM_{\rm PV})^3$ and $\mu_{\rm A}\propto \rho_{\rm A}(k/aM_{\rm PV})^3$. 
{\color{black}Again, we find these terms would be more suppressed by the energy scale $M_{\rm PV}$ in the view of effective field theories.}

We observe that in all these theories, except for CS modified gravity, the velocities of left-hand and right-hand polarizations are different, {\emph{i.e.}}, the effect of velocity birefringence exists.

\section{Reconstruction of the complex waveform of GWs from the observations \label{appendixB}}

In the real observations of GW detectors, the observables are the {\emph{real}} parts of $h_{+}(t)$ and $h_{\times} (t)$. We consider $h_+$ as an example, which can be written as the following (general) form
\begin{equation}
h_+(t)= A(t_{r})e^{-i\Phi(t_{r})},
\end{equation}
and the real part of this wave is
\begin{equation}
h_+^{\mathfrak{R}}(t)\equiv{\mathfrak R}\{ h_+(t)\}=A(t_{r})\cos\Phi(t_{r}),
\end{equation}
where $t_r$ is the retarded time. The terms $A$ and $\Phi$ represent the amplitude and phase of GWs respectively, which are both functions of $t_r$.

For GWs produced during the inspiralling stage of compact binaries, the conditions $d\ln A/dt\ll d\Phi/dt$ and $|d^2\Phi/dt^2|\ll(d\Phi/dt)^2$ are satisfied. Therefore, we can use SPA to obtain the waveform in Fourier domain $\tilde{h}_{+}^{\mathfrak R}(f)$ as follows \cite{spa},
\begin{eqnarray}
&&\tilde{h}_{+}^{\mathfrak R}(f)=\int dt A(t_{r}) \cos\Phi(t_{r}) e^{i2\pi ft} \nonumber \\
&&=\frac{1}{2}e^{i2\pi fr/c}\int dt_{r} A(t_{r})\left(e^{i\Phi(t_{r})}+e^{-i\Phi(t_{r})}\right)e^{i2\pi f t_{r}} \nonumber\\
&&=\frac{1}{2} e^{i\Psi_+}A(t_*) \left(\frac{2\pi}{\ddot{\Phi}(t_*)}\right)^{1/2}~~~(f>0),\label{a1}
\end{eqnarray}
and
\begin{equation}
\tilde{h}_{+}^{\mathfrak R}(-f)=(\tilde{h}_+^{\mathfrak R})^*(f) ~~~(f>0),\label{a2}
\end{equation}
where $\Psi_+(t) \equiv 2\pi f t_r -\Phi(t_r)-\pi/4$. In this equation $t_*$ is defined as the time at which $d\Phi/dt=2\pi f$, and $\Psi_+(t_*)$ is the value of $\Psi_+(t)$ at $t=t_*$.

Similarly, assuming SPA is applicable, we can also obtain the $h_+(t)$ in Fourier domain $\tilde{h}_{+}(f)$ as follows,
\begin{eqnarray}
&&\tilde{h}_{+}(f)=\int dt A(t_{r}) e^{-i\Phi(t_{r})} e^{i2\pi ft}\\
&&=e^{i\Psi_+}A(t_*) \left(\frac{2\pi}{\ddot{\Phi}(t_*)}\right)^{1/2}~~~(f>0),\label{b1}
\end{eqnarray}
and
\begin{equation}
\tilde{h}_{+}(-f)=0 ~~~(f>0). \label{b2}
\end{equation}

By comparing Eqs.(\ref{a1}), (\ref{a2}), (\ref{b1}), (\ref{b2}), we obtain the following relation:
\bea\label{c2}
\tilde{h}_{+}=\left\{
 {\begin{array}{l}
 2\tilde{h}_{+}^{\mathfrak{R}},~f>0,  \\
 0, ~~~~~f<0,
 \end{array}
 }
 \right.
\ena
which indicates that one can obtain the Fourier component $\tilde{h}_{+}$ from $\tilde{h}_{+}^{\mathfrak{R}}$. However, we should emphasize that, this relation is applicable only to the GW produced during the inspiraling stage of compact binaries, where SPA is correct. 

Once we obtain $\tilde{h}_+$ and $\tilde{h}_{\times}$, the Fourier components of $h_{\rm R}$ and $h_{\rm L}$ are derived straightforwardly,
\begin{eqnarray}\label{e1}
\tilde{h}_{\rm R}&=&\frac{1}{\sqrt{2}}\left[\tilde{h}_+-i\tilde{h}_{\times}\right], \nonumber \\ \tilde{h}_{\rm L}&=&\frac{1}{\sqrt{2}}\left[\tilde{h}_++i\tilde{h}_{\times}\right].
\end{eqnarray}
Using the inverse Fourier transformation, we can also obtain the time-domain functions $h_{\rm R}(t)$ and $h_{\rm L}(t)$.

\section{Decomposition of the circular polarization modes of GW170817 \label{appendixC}}

We apply the analysis described above to GW170817, which is the first and loudest GW burst with observed EM counterpart \cite{gw170817}. We use the strain data after noise subtraction of AdvLIGO detectors, and adopt the data with total duration of 2048 seconds and sampling frequency 4096Hz. Note that, we do not use the AdvVirgo data, in which the SNR of GW signal is too weak. The parameters of GW source are given by the EM counterpart: The position is at (RA=$13^{\rm h}09^{\rm m}48.08^{\rm s}$, Dec=$-23^{\circ}22'53.3''$), and the time of merger is 12:41:04 UTC, 17August 2017 (GPS time 1187008882.43$s$) \cite{gw170817}. The polarization angle can be randomly chosen as mentioned above, so we adopt it as $\psi_{s}=0^{\circ}$. Before analysis, we filter all time serials with a 20-500Hz bandpass filter to suppress large fluctuations outside the detector's most sensitive frequency band.

We repeat the reconstruction method as before, but here the mock data is replaced with the real data of LIGO detectors. Following the decomposition proceeding, we first construct the unbiased estimators for $\tilde{h}_{+}^{\mathfrak{R}}$ and $\tilde{h}_{\times}^{\mathfrak{R}}$, and translate them to $\tilde{h}_{+}$ and $\tilde{h}_{\times}$, respectively. Based on these, the left-hand and right-hand polarizations in frequency domain and time domain are derived straightforwardly.

We show the results in the time-frequency representation, which are given in the upper panels of Figure 3. From both diagrams, we find the GW signal is weak (As expected, they are even weaker than those in simulations), and it seems difficult to identify them directly by eyes. Therefore, we should use the frequency superimposition technique to amplify signal and get the arrival times of both GW polarizations, where we adopted $\mathcal{M}_c=1.188{\rm M}_{\odot}$ derived in GR framework to approximately displace the true value of chirp mass \cite{gw170817}. In the analysis, we scan the arrival times of the frequency bins with 5Hz in length from $(20,25)$Hz, $(25,30)$Hz to $(195,200)$Hz. For both polarizations, we observe the maximum SNR at the bin of $(140,145)$Hz. Similarly, we change the frequency bins with 10Hz in length, and find the best one is at $(135,145)$Hz. The arrival times of circular polarizations are presented in Figure 3 (low panels).

For the arrival time of right-hand polarization, we observe the significant peaks ($\gtrsim 4\sigma$) for both frequency bins (140,145)Hz and (135,145)Hz, which follows the consistent results, {\emph{i.e.}}, $t_{\rm R}=-0.911^{+0.060}_{-0.062}s$ for the former bin, and $t_{\rm R}=-0.901^{+0.063}_{-0.063}s$ for the latter bin. Note that, for each bin, we have calibrated the arrival times of all the frequency bands in the bin to a fixed frequency at 140Hz. Since here we have used the blind method for the signal search, and consider the signal only in a small frequency bin, the SNR is much lower than that derived from template fitting ($\sim 32 \sigma$). On the other hand, the SNR of arrival time for left-hand polarization is too low for both frequency bins. These results are consistent with what we anticipate: Since the inclination angle of GW170817 is $\iota\ge152^{\circ}$ \cite{gw170817}, the amplitude ratio of circular polarization modes $|h_{\rm L}|/|h_{\rm R}|<0.004$. Therefore, the observed GW signal of this event is completely dominated by the right-hand mode.

\begin{figure}[htbp]
\small
\centering
\includegraphics[width=8.5cm]{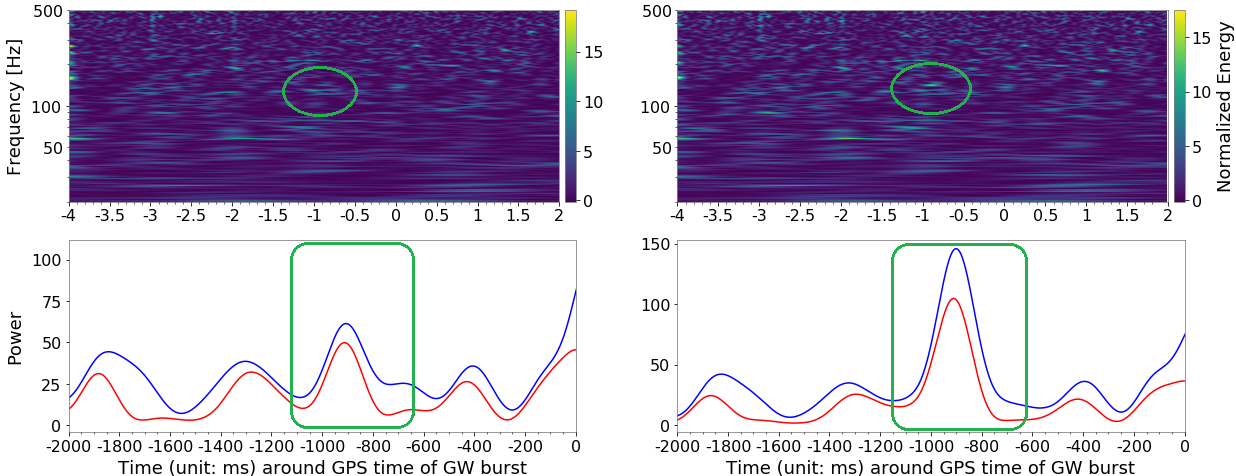}
\caption{ The left-hand (left panels) and right-hand (right panels) polarizations of GW event GW170817. Times are shown relative to GPS time 1187008882.43. Upper panels show the time-frequency representations, and lower panels show the arrive times (unit: ms), where we combine the signal during (135,145)Hz for blue curves and (140,145)Hz for red curves.}
\label{fig4}
\end{figure}


\begin{thebibliography}{99}
\bibitem{ft0}
Note that, the requirement on the inclination angle $\iota$ is needed only for the method introduced in this paper. If we constrain the parity symmetry of gravity by the template fitting of the GW waveform, this condition is not necessary \cite{wangyifan,huqian}.
\bibitem{ft1}
Chern-Simons gravity is a special theory of parity-violating gravity, as demonstrated in Appendix A, in which only amplitude birefringence effect exists. Therefore, the constraint derived above in not applicable.
\end{thebibliography}
\end{document}